\documentclass[journal]{IEEEtran}
\usepackage[letterpaper,margin=1in]{geometry}
\usepackage{amsmath,amssymb,amsthm}
\usepackage{graphicx}
\usepackage{cite}
\usepackage{booktabs}
\usepackage[hidelinks]{hyperref}

\newtheorem{theorem}{Theorem}
\newtheorem{lemma}{Lemma}
\newtheorem{proposition}{Proposition}
\newtheorem{corollary}{Corollary}
\theoremstyle{remark}

\newcommand{\E}{\mathbb{E}}
\newcommand{\lam}{\lambda}
\newcommand{\cH}{\mathcal{H}}
\newcommand{\cK}{\mathcal{K}}
\newcommand{\Ae}{A_{\mathrm e}}

\begin{document}

\title{Support Transitions of the Amplitude-Constrained Gaussian Channel: Location, Splitting Law, and Second-Order Capacity Behavior}

\author{Gang~Liu, Guosheng~Xu, and~Guoai~Xu%
\thanks{(Corresponding author: Guosheng Xu.)}%
\thanks{Gang Liu and Guosheng Xu are with the School of Cyberspace Security, Beijing University of Posts and Telecommunications, Beijing 100876, China (e-mail: gang.liu@bupt.edu.cn; guoshengxu@bupt.edu.cn). Gang Liu is also with Trans-Asia Gas Pipeline Company Limited, Beijing, China.}%
\thanks{Guoai Xu is with the School of Computer Science and Technology, Harbin Institute of Technology, Shenzhen 518055, China (e-mail: xga@hit.edu.cn).}%
\thanks{Code and certificate files: Zenodo, version~1.1.0, doi:~10.5281/zenodo.24641857.}}

\ifCLASSOPTIONonecolumn
\IEEEaftertitletext{\vspace{-\baselineskip}}
\fi
\maketitle
{\footnotesize\noindent This work has been submitted to the IEEE for possible publication. Copyright may be transferred without notice, after which this version may no longer be accessible.\par}

\begin{abstract}
We characterize the first three support transitions of the amplitude-constrained scalar Gaussian channel and the first two of its quadrature counterpart. At each transition, the rate loss from retaining the previous constellation family is quadratic in the excess amplitude, with a strictly positive coefficient. A new mass grows linearly from zero; when the mass at the origin splits, the inner radius grows as the square root of the excess amplitude. In the scalar channel the binary input remains optimal through $A_1\approx1.666$. The ternary input is optimal on $(A_1,A_2]$ and the four-point input on $(A_2,A_3]$, where $A_2\approx2.9074640719$ and $A_3\approx4.02127532$. We enclose $A_2$ in an interval narrower than $10^{-10}$, confirming the numerical estimate $2.9075$ rather than the previously reported $2.786$. At $A_1$ and $A_3$ the origin meets the optimality condition with equality without carrying mass; it acquires positive mass just to the right of each threshold. Local expansions of mutual information and interval verification of global optimality show that capacity is continuously differentiable but not twice differentiable at all five transitions.
\end{abstract}

\begin{IEEEkeywords}
Amplitude constraints, ball arithmetic, channel capacity, Gaussian channels, support transitions.
\end{IEEEkeywords}

\section{Introduction}

\IEEEPARstart{C}{onsider} the scalar channel $Y=X+Z$, with $Z\sim\mathcal N(0,1)$ independent of $X$, $|X|\le A$, and capacity $C(A)=\max I(X;Y)$. Its capacity-achieving input distribution (CAID) is unique and symmetric, with finite support containing $\pm A$~\cite{smith1971}. How does that support change as $A$ increases, and what rate is lost by retaining the previous constellation? Discreteness extends to certain conditionally Gaussian channels~\cite{chan2005}; in the quadrature and isotropic vector channels the optimal input has finitely many radii of support~\cite{shamai1995,rassouli2016}. In the scalar case the support size grows at most quadratically~\cite{dytso2020} and faster than linearly~\cite{wang2025a,wang2025}. Related bounds concern near-optimal inputs~\cite{barletta2026eps} and Poisson channels~\cite{barletta2024poisson}. These bounds do not locate individual transitions.

\subsection{Prior Work}
Sharma and Shamai~\cite{sharma2010} derived necessary conditions for the appearance and splitting of a mass at the origin. Their sufficient conditions assumed that the support gains at most one point at a time; they reported a ternary-to-four-point transition at $A\approx2.786$. Without that assumption, Dytso et al.~\cite[Thm.~2]{dytso2019} characterized exactly when a single sphere is optimal in any dimension. Their scalar estimate $\bar R_1\approx1.665925641$~\cite[Remark~2]{dytso2019} lies in our certified bracket.

Barletta and Dytso~\cite{barletta2026} proved that local changes in support cardinality occur only at the origin, by appearance, disappearance, splitting, or merging; the change is at most one. They reported a numerical estimate of $2.9075$ for the ternary-to-four-point transition~\cite[footnote~2]{barletta2026}. Ikeda et al.~\cite{ikeda2014} numerically traced the optimal locations for box and circular constraints. Using a dynamic-assignment Blahut--Arimoto algorithm, Xiao et al.~\cite[Fig.~1]{xiao2021} observed the scalar split at a peak signal-to-noise ratio of $9.28$~dB ($10\log_{10}A^2$ in our normalization). These computations locate the split numerically; they do not certify its location or determine the rate loss from retaining the ternary input. A different transition, between discrete and continuous inputs, occurs for additive uniform noise~\cite{stapmanns2026}.

\subsection{Contributions}
At the five transitions studied here the rate loss is quadratic, even when the new support radius grows as a square root. Theorem~\ref{thm:main} encloses the scalar transition amplitude $A_2$ in an interval of width less than $10^{-10}$. The crossing is transversal: if $g(A)$ denotes the slope at the origin of the optimized ternary input's information density as a function of squared radius, then $g(A_2)=0$ and $g'(A_2)>0$. Fig.~\ref{fig:support} traces the scalar support through the third transition. The ternary input is optimal throughout $(A_1,A_2]$, including at $2.786$; the four-point input is optimal throughout $(A_2,A_3]$ (Theorems~\ref{thm:main} and~\ref{thm:third}). Fig.~\ref{fig:kkt} illustrates why the ternary input remains optimal at the earlier reported amplitude but fails at $A=3$. At $A_1$ and $A_3$ the optimality inequality is tight at the origin although the origin is not in the support. Just above either threshold, the origin carries positive mass. Theorem~\ref{thm:quad} identifies the first two analogous transitions for the quadrature channel, from one circle to a circle and the origin, then to two positive radii.

Write $\delta=A-A_i>0$. A newly appearing mass is $\ell_i\delta+O(\delta^2)$; after a split, the squared inner radius is $c\delta+O(\delta^2)$. In all five cases the loss from retaining the previous constellation family is $K_i\delta^2+O(\delta^3)$ with $K_i>0$, so that the first derivative of capacity is continuous and the second derivative jumps by $2K_i$. The comparison allows the old constellation to readjust at the new amplitude: the ternary weight is reoptimized above $A_2$, and the four-point input follows its stationary branch above $A_3$. Propositions~\ref{prop:appear}, \ref{prop:split}, and~\ref{prop:reentry} give the local laws under explicit nondegeneracy conditions; Theorems~\ref{thm:laws}--\ref{thm:quad} certify the conditions and coefficients.

\subsection{Method and Outline}
We reduce global optimality to Karush--Kuhn--Tucker (KKT) inequalities in the squared input radius. The certificates~\cite{p2pcert} verify these inequalities on entire cells using ball arithmetic~\cite{arb,johansson2018} and enclose stationary solutions with parametric Krawczyk tests~\cite{krawczyk1969,moore1977}. Near $A_2$ and $A_3$ we use the squared inner radius or the new mass as the parameter so that the certificates reach the threshold itself.

Section~\ref{sec:kkt} gives the optimality conditions. Sections~\ref{sec:results} and~\ref{sec:cert} establish the first two scalar transitions; Sections~\ref{sec:laws} and~\ref{sec:third} derive the local laws and treat the third transition. Section~\ref{sec:quad} gives the quadrature results and Section~\ref{sec:disc} discusses the scope of the certificates. Appendices~\ref{app:sound} and~\ref{app:formulas} give the cell proofs and integration bounds.

\section{Optimality Conditions}
\label{sec:kkt}

\subsection{KKT Conditions and the Binary Regime}

We use natural logarithms: $I$, $C$, and the coefficients $K_i$ in the rate expansions are in nats. For rates reported in bits, divide by $\ln 2$; in particular, write $K_i^{\rm bit}=K_i/\ln 2$. The mass and radius coefficients $\ell_i$ and $c$ are unchanged.
Let $\varphi$ be the standard normal density, $f_P=\varphi*P$, and $i(x;P)=D(\mathcal N(x,1)\|f_P)$, so that $I(P)=\int i(x;P)\,dP(x)$. The following is standard~\cite{smith1971,huang2005}.

\begin{lemma}\label{lem:kkt}
$I$ is strictly concave on the distributions on $[-A,A]$, and $P$ is capacity-achieving if and only if $i(x;P)\le I(P)$ for all $|x|\le A$, with equality on the support of $P$. Moreover, $I(P')-I(P)=\int i(x;P)\,d(P'-P)-D(f_{P'}\|f_P)$ for all $P,P'$.
\end{lemma}

For symmetric $P$, $i(x;P)$ depends on $s=x^2$; write $\Phi_P(s)=i(\sqrt s;P)$. Let $B_A=\frac12(\delta_{-A}+\delta_A)$ and
\begin{equation}
P_\lam=\lam\delta_0+(1-\lam)B_A,\qquad 0\le\lam<1 .
\label{eq:Plam}
\end{equation}
Then $f_{P_\lam}(y)=\varphi(y)M_\lam(y^2)$ with $M_\lam(q)=\lam+(1-\lam)e^{-A^2/2}\cosh(A\sqrt q)$, and, with $Q_{m,s}=\|\sqrt s\,e_1+Z_m\|^2$, $Z_m\sim\mathcal N(0,I_m)$,
\begin{equation}
\Phi_\lam(s):=\Phi_{P_\lam}(s)=\frac s2-\E\ln M_\lam(Q_{1,s}).
\label{eq:Phi}
\end{equation}
Since $\partial_s\E f(Q_{m,s})=\E f'(Q_{m+2,s})$ (the law of $Q_{m,s}$ is a $\mathrm{Poisson}(s/2)$ mixture of the laws $\chi^2_{m+2j}$; differentiate the weights and use $\E f(\chi^2_{k+2})-\E f(\chi^2_k)=2\E f'(\chi^2_{k+2})$), $\Phi_\lam'(s)=\frac12-\E(\ln M_\lam)'(Q_{3,s})$ and $\Phi_\lam''(s)=-\E(\ln M_\lam)''(Q_{5,s})$. By Lemma~\ref{lem:kkt},
\begin{equation}
\cH(\lam,A):=\partial_\lam I(P_\lam)=\Phi_\lam(0)-\Phi_\lam(A^2),
\label{eq:H}
\end{equation}
and $\cK_\lam:=-\partial_\lam\cH=\E[\psi(Y)^2]>0$ with $\psi=(f_{\delta_0}-f_{B_A})/f_{P_\lam}$ and $Y\sim f_{P_\lam}$; thus $\cH(\cdot,A)$ is strictly decreasing.

\begin{proposition}[Binary optimality]\label{prop:binary}
For every $A>0$, $\Phi_0$ is strictly convex on $[0,\infty)$:
\begin{equation*}
\Phi_0''(s)=\int_0^{1/2}u\,\beta(u)\,e^{-us}\,du>0,
\end{equation*}
with $\beta(u)=(1-2u)^{-1/2}\,\Theta\big(\frac{u}{A^2(1-2u)}\big)$ and $\Theta(t)=\sum_{k\ge1}e^{-(k-\frac12)^2\pi^2t}$. Hence $B_A$ is capacity-achieving if and only if $\cH_0(A):=\cH(0,A)\le0$. The function $2\cH_0(A)/A^2$ is continuous and strictly increasing with limits $\mp1$, so there is a unique $A_1$ with $\cH_0<0$ on $(0,A_1)$ and $\cH_0>0$ on $(A_1,\infty)$.
\end{proposition}
\begin{proof}
From $\cosh z=\prod_{k\ge1}(1+z^2/((k-\frac12)^2\pi^2))$~\cite[\S4.36]{dlmf}, we obtain
\begin{multline*}
\bigl(\ln\cosh(A\sqrt q)\bigr)'
 = \sum_k \left(\frac{(k-1/2)^2\pi^2}{A^2}+q\right)^{-1}\\
 = \int_0^\infty \Theta(t/A^2)e^{-tq}\,dt .
\end{multline*}
With $\E e^{-tQ_{3,s}}=(1+2t)^{-3/2}e^{-st/(1+2t)}$ and the substitution $u=t/(1+2t)$, $\E(\ln M_0)'(Q_{3,s})=\int_0^{1/2}\beta(u)e^{-us}du$, which gives $\Phi_0'$ and, after differentiation, $\Phi_0''$. A strictly convex $\Phi_0$ attains its maximum over $[0,A^2]$ at an end point, and $\Phi_0(A^2)=I(B_A)$, so Lemma~\ref{lem:kkt} gives the first claim. For the second claim, let $\hat x_v(y)=A\tanh(Ay/v)$ be the posterior mean under $B_A$ at noise variance $v$. At true input zero the squared error has mean $\E[\hat x_v(\sqrt vZ)^2]$. Averaged over the two outer inputs it equals the binary minimum mean-square error, $A^2-\E[\hat x_v(A+\sqrt vZ)^2]$, by orthogonality and symmetry. Let $Q_v$ be the output law under $B_A$ at noise variance $v$. By convexity of relative entropy in its second argument, $0\le D(\mathcal N(x,v)\|Q_v)\le (x^2+A^2)/(2v)$, the upper bound being the average of the divergences to $\mathcal N(-A,v)$ and $\mathcal N(A,v)$. Thus the boundary terms at $x=0,\pm A$ vanish as $v\to\infty$. Integrating the relative de~Bruijn identity~\cite{verdu2010} for the difference of the two errors gives
\begin{multline*}
\frac{2\cH_0(A)}{A^2}
 = \int_1^\infty \biggl[
   \E\tanh^2\!\left(\frac{A|Z|}{\sqrt v}\right)\\
   + \E\tanh^2\!\left(\frac{A|A+\sqrt vZ|}{v}\right) - 1
   \biggr]\frac{dv}{v^2},
\end{multline*}
which also gives the limits: dominated convergence applies because the bracket is bounded by $1$ in absolute value. For fixed $v$, $|A+\sqrt vZ|$ stochastically increases with $A\ge0$: for $t>0$ the derivative of $\Pr(|A+\sqrt vZ|\le t)$ is $\big[\varphi((t+A)/\sqrt v)-\varphi((t-A)/\sqrt v)\big]/\sqrt v\le0$. Multiplication by $A$ preserves this order, and $\tanh^2$ is increasing on the positive half-line. The first expectation increases strictly, proving the claimed monotonicity.
\end{proof}

The optimality threshold in Proposition~\ref{prop:binary} is the scalar case of~\cite[Thm.~2]{dytso2019}. The kernel representation gives an alternative proof that also applies in Section~\ref{sec:quad}. Since $\cH(1,A)=-A^2/2$ by continuity at the pure zero input, for $A>A_1$ $\cH(\cdot,A)$ has a unique root $\lam^*(A)\in(0,1)$, the maximizer of $I(P_\lam)$ over $\lam$. The next statement is the splitting condition of~\cite[eq.~(40)]{sharma2010} in the variable $s=x^2$: since $\frac{d^2}{dx^2}\Phi(x^2)|_{x=0}=2\Phi'(0)$, it says that the information density must not be convex at the origin.

\begin{proposition}[Necessary condition for ternary optimality]\label{prop:exit}
If $A>A_1$ and some $P_\lam$, $\lam\in(0,1)$, is capacity-achieving, then $\lam=\lam^*(A)$ and $g(A):=\Phi'_{\lam^*(A)}(0)\le0$.
\end{proposition}
\begin{proof}
Lemma~\ref{lem:kkt} forces $\Phi_\lam(0)=\Phi_\lam(A^2)=I(P_\lam)$, i.e., $\cH(\lam,A)=0$, and $\Phi_\lam(s)\le\Phi_\lam(0)$ for small $s$.
\end{proof}

Let $A_2=\sup\{A'>A_1:\ P_{\lam^*(A)}\text{ is optimal for all }A\in(A_1,A']\}$ and $\Ae=\inf\{A>A_1:\ g(A)>0\}$; Proposition~\ref{prop:exit} gives $A_2\le\Ae$.

\subsection{A Sufficient Condition Near the First Transition}\label{sec:mixgate}
We give a single argument for the scalar and quadrature cases. Write $n=1$ or $2$, let $U_A$ be the uniform law on the sphere of radius $A$ ($U_A=B_A$ for $n=1$), and consider $P_\lam=\lam\delta_0+(1-\lam)U_A$. With $q=\|y\|^2$, write $M_0(q)=e^{-A^2/2}L(q)$, where $L(q)=\cosh(A\sqrt q)$ or $I_0(A\sqrt q)$, respectively. In dimension $n$, \eqref{eq:Phi} reads $\Phi_\lam(s)=s/2-\E\ln M_\lam(Q_{n,s})$, with $M_\lam=\lam+(1-\lam)M_0$; \eqref{eq:H} and the information-density criterion follow from the same output-density identity in Lemma~\ref{lem:kkt}. Put $b(q)=1/M_0(q)$ and define
\begin{align*}
m&=\Phi_0''(A^2),\qquad \alpha=\E b''(Q_{n+4,0}),\\
\cK&=\cK_0=\chi^2(f_{\delta_0}\|f_{U_A}),\\
T&=\max\{1,e^{A^2/2}-1\}\cK,\\
\lam_0&=\min\{1/2,\ m/(m+2\alpha),\ \cK/(\cK+2T)\}.
\end{align*}
\begin{lemma}[sufficient condition]\label{lem:gate}
Let $n\in\{1,2\}$ and $A>0$. If $0<\cH_0(A)<\lam_0\cK/2$, then $\cH(\lam,A)=0$ has a unique root $\lam^*\in(0,\lam_0)$ and $P_{\lam^*}$ is capacity-achieving.
\end{lemma}

\begin{proof}
In both dimensions $L(q)=\prod_{j\ge1}(1+q/d_j)$ with $\sum_j d_j^{-1}<\infty$. For $n=1$, $d_j=((j-1/2)\pi/A)^2$; for $n=2$, $d_j=(j_{0,j}/A)^2$, where $j_{0,j}$ is the $j$th positive zero of $J_0$~\cite[\S\S4.36, 10.21, 10.27]{dlmf}. Thus $-(\ln M_0)''(q)=\sum_j(d_j+q)^{-2}$ is positive and decreasing. Also $e^{-A^2/2}b$ is the Laplace transform of the sum of independent exponentials of means $1/d_j$; hence $b''$ is nonnegative and decreasing (its value at zero is finite). The Poisson-mixture formula above shows that $Q_{n+4,s}$ increases stochastically with $s$. Applying the radial derivative identity twice with initial dimension $n$ gives $\Phi_0''(s)\ge m>0$ for $0\le s\le A^2$.

Write $M_\lam=M_0(1-\lam+\lam b)$ and $\ell=\ln(1-\lam+\lam b)$. Differentiation gives $\ell''\le\lam b''/(1-\lam)$; hence $\Phi_\lam''(s)\ge m-\lam\alpha/(1-\lam)$. For $\beta=b-1$, the output-density formula gives $\cK_\lam=\E_{f_{U_A}}[\beta^2/(1+\lam\beta)]$, with $\beta$ evaluated at $\|Y\|^2$. Since $-1\le\beta\le e^{A^2/2}-1$, we have $|\cK_\lam-\cK|\le\lam T/(1-\lam)$. For $0\le\lam\le\lam_0$ these inequalities give $\Phi_\lam''\ge m/2$ and $\cK_\lam\ge\cK/2$. Integrating $\partial_\lam\cH=-\cK_\lam$ now gives $\cH(\lam_0,A)<0<\cH(0,A)$. At its unique root the endpoint densities agree and equal $I(P_{\lam^*})$. Strict convexity puts every interior value below that common value, proving the global optimality inequality.
\end{proof}

\section{The First Two Scalar Transitions}
\label{sec:results}
Write $P_{\mu,\rho}=\frac{\mu}{2}(\delta_{-\sqrt\rho}+\delta_{\sqrt\rho})+(1-\mu)B_A$ for the symmetric four-point input. The following result identifies the input on either side of the split; its proof follows the certificate construction in Section~\ref{sec:cert}. Tables~\ref{tab:tern} and~\ref{tab:four} give the corresponding probabilities, radii, and rates at selected amplitudes.

\begin{theorem}\label{thm:main}
The first two scalar transition points satisfy $A_1\in[1.6659256408,1.6659256470]$ and $A_2\in[2.907464071846,2.907464071942]$.
\begin{enumerate}
\item[(i)] The binary input $B_A$ is capacity-achieving if and only if $A\le A_1$.
\item[(ii)] For every $A\in(A_1,A_2]$, the CAID is the ternary input $P_{\lam^*(A)}$ of~\eqref{eq:Plam}, with $0<\lam^*(A)<1$.
\item[(iii)] For every $A\in(A_2,3.98]$, no input with at most three mass points is capacity-achieving.
\item[(iv)] The function $g(A)=\Phi'_{\lam^*(A)}(0)$ is negative on $(A_1,2.9074]$, strictly increasing on $[2.9074,2.9075]$ with $g'\in[0.1660,0.1719]$, and positive on $[2.9075,3.2]$. Hence $A_2=\Ae$ is a transversal zero of $g$.
\item[(v)] For every $A\in(A_2,3.98]$, the CAID is a four-point input $P_{\mu^*,\rho^*}$ with $0<\rho^*<A^2$. In particular, the origin does not belong to its support.
\end{enumerate}
\end{theorem}

\begin{table*}[t]
\centering
\caption{Certified values in the ternary regime: bracket of $\lam^*(A)$ (rounded outward), capacity $C(A)$ in bits (error below $10^{-10}$), the rate of the binary input $B_A$, and the relative gain $C/I(B_A)-1$ (lower bound).}
\label{tab:tern}
\setlength{\tabcolsep}{8pt}
\begin{tabular}{ccccc}
\toprule
$A$ & $\lam^*(A)$ & $C(A)$ & $I(B_A)$ & gain\\
\midrule
$1.8$ & $[0.066104947,0.066104948]$ & $0.8704182473$ & $0.8653688342$ & $0.58\%$\\
$2.0$ & $[0.137876946,0.137876947]$ & $0.9418905971$ & $0.9128222858$ & $3.18\%$\\
$2.5$ & $[0.237250014,0.237250015]$ & $1.1181186842$ & $0.9751790043$ & $14.65\%$\\
$2.786$ & $[0.267101063,0.267101064]$ & $1.2095794127$ & $0.9891442855$ & $22.28\%$\\
$2.9074$ & $[0.276496430,0.276496431]$ & $1.2452374872$ & $0.9925378260$ & $25.45\%$\\
\bottomrule
\end{tabular}
\end{table*}

\begin{table}[t]
\centering
\caption{Four-point inputs $\{\pm r^*,\pm A\}$ (Theorem~\ref{thm:main}(v)): total inner mass $\mu^*$ and inner radius $r^*$ (errors below $2\times10^{-6}$), certified capacity in bits, and the floating-point rate of the best ternary input $P_{\lam}$, $\lam\in[0,1)$.}
\label{tab:four}
\small
\setlength{\tabcolsep}{4pt}
\begin{tabular}{@{}ccccc@{}}
\toprule
$A$ & $\mu^*$ & $r^*$ & $C(A)$ & best ternary\\
\midrule
$3.0$ & $0.294030$ & $0.394728$ & $[1.27153,1.27154]$ & $1.27103$\\
$3.2$ & $0.327613$ & $0.681082$ & $[1.32753,1.32754]$ & $1.32239$\\
$3.5$ & $0.367883$ & $0.934949$ & $[1.40907,1.40908]$ & $1.38813$\\
$3.9$ & $0.406570$ & $1.169223$ & $[1.51151,1.51153]$ & $1.45573$\\
\bottomrule
\end{tabular}
\end{table}

\begin{corollary}\label{cor:2786}
At $A=2.786$ the ternary input is capacity-achieving, with $\lam^*\in[0.267101063,0.267101064]$ and $g(2.786)=-0.0196048\pm10^{-7}$. Hence $2.786$ is not a transition point, and the first regime of optimality of the ternary input ends at $A_2\in[2.9074640718,2.9074640720]$, where the four-point input takes over.
\end{corollary}
\begin{proof}
Theorem~\ref{thm:main}(ii) gives optimality. At $A=2.786$, the mass bracket in Table~\ref{tab:tern} and the pointwise enclosure of $\Phi'_{\lam^*}(0)$ in~\cite{p2pcert} give the stated numerical bounds.
\end{proof}

Theorem~\ref{thm:third} in Section~\ref{sec:third} extends (v) up to the third transition $A_3\approx4.02128$. Capacities in the ternary regime use the concavity of $\lam\mapsto I(P_\lam)$: $I(P_{\lam_-})\le C\le I(P_{\lam_-})+\cH(\lam_-,A)(\lam_+-\lam_-)$ for a bracket $[\lam_-,\lam_+]$ of $\lam^*$ of width below $10^{-11}$; in the four-point regime, $C$ is enclosed by evaluating $I$ over the Krawczyk box. At $A=2.9074$, just below $A_2$, the zero symbol raises the rate over the binary input by at least $25.45\%$. In the four-point regime, the gain over the best ternary input is small at first, $0.04\%$ at $A=3.0$, and $3.8\%$ at $A=3.9$ (floating point).

\section{Certificates}
\label{sec:cert}

\subsection{Ball Arithmetic}
All quantities above are one-dimensional integrals in $u=y^2/2$ against Gamma-type densities, e.g., $u$ has density $e^{-u-z}u^{-1/2}\,{}_0F_1(;\frac12;zu)/\Gamma(\frac12)$ under $\mathcal N(\sqrt s,1)$, $z=s/2$. We enclose them with the ball-arithmetic integrator of Arb~\cite{arb,johansson2018} (FLINT~3.6~\cite{flint}, python-flint~0.9.0, $100$ bits) in the variable $t=\sqrt u\in[0,13]$, with explicit truncation bounds. Appendix~\ref{app:formulas} gives the tail bounds for the zero--sphere mixture, the four- and five-point inputs, and their parameter derivatives at unit noise variance. The certificate implementation is archived in~\cite{p2pcert}. Logarithms and reciprocals are evaluated in Arb's analytic mode, so that a ball that may meet a branch cut or a pole is rejected and subdivided. All acceptance tests compare balls; floating point is used only to propose brackets and subdivisions.

\subsection{Cells}
For a parameter cell $[A_l,A_r]$ we first enclose $\lam^*(A)$: with $\Lambda=[\lam_-,\lam_+]$ proposed in floating point, we accept if $\cH(\lam_-,A_m)-|\partial_A\cH|\delta>0>\cH(\lam_+,A_m)+|\partial_A\cH|\delta$, $A_m$ the midpoint and $\delta$ the half-width; then $\lam^*(A)\in\Lambda$ on the cell. With $G=\Phi_\lam-\Phi_\lam(0)$, a cell $C=[t_i,t_{i+1}]$ in $s$ is of type D if $G'<0$ on $C$, C if $G''<0$ on $C$ and $G'(t_i)<0$, I if $G'>0$, V if $G<0$ (for this $G$, $G(0)=0$ forces $t_i>0$), and C$_0$ if $G''<0$, each enclosed over $[A_l,A_r]\times\Lambda\times C$.

\begin{lemma}\label{lem:sound}
Let $0=t_0<\dots<t_m=A_r^2$. Suppose that the cells $C_i$, $i<k_0$, are of type C$_0$, those with $k_0\le i<k$ of type D, C, or V, and those with $i\ge k$ of type I or V, every cell with $t_{i+1}>A_l^2$ being of type I. Then, for every $A\in[A_l,A_r]$, $P_{\lam^*(A)}$ is capacity-achieving if and only if $g(A)\le0$. If $k_0=0$, $P_{\lam^*(A)}$ is capacity-achieving for every $A$ in the cell.
\end{lemma}

Appendix~\ref{app:sound} proves the lemma by induction from $G(0)=G(A^2)=0$. We use the following three algorithms on a cell $[A_l,A_r]$ with the certified mass bracket $\Lambda$ above. Each bound is required on its entire cell. A failed test gives no conclusion; a proof of coverage requires accepted cells with no gaps.

\emph{Algorithm~1 (optimality).} Partition $[0,A_r^2]$ and accept if the D/C/V and I/V tests of Lemma~\ref{lem:sound} succeed with $k_0=0$, including the type-I requirement at the moving right endpoint. Then $G\le0$ for every admissible radius and every $A$ in the parameter cell.

\emph{Algorithm~2 (exclusion).} Enclose $\Phi'_\lam(0)$ on $[A_l,A_r]\times\Lambda$. A strictly positive lower bound implies $g(A)>0$ throughout the cell, so Proposition~\ref{prop:exit} excludes a ternary optimum.

\emph{Algorithm~3 (conditional optimality).} Verify $G''<0$ from $0$ to $t_{k_0}=0.05$ and apply the remaining cell tests of Lemma~\ref{lem:sound} from that endpoint onward. The verdict is now conditional: the input is optimal exactly where $g(A)\le0$, including equality. We certify the sign on subcells away from the crossing and use the next derivative test on the cell containing the crossing.

\subsection{Transversality}
On the gap cell, Lemma~\ref{lem:sound} reduces optimality to the sign of $g$. We enclose
\begin{equation}
g'(A)=\partial_A\Phi'_\lam(0)+\partial_\lam\Phi'_\lam(0)\,\frac{\partial_A\cH(\lam,A)}{\cK_\lam}\Big|_{\lam=\lam^*(A)}
\label{eq:gprime}
\end{equation}
over the cell times $\Lambda$, with the formulas of Appendix~\ref{app:formulas}. If the enclosure is positive, $g$ has at most one zero on the cell, and certified signs of $g$ at rational points, each from a bracket of $\lam^*$ refined by certified bisection to width $10^{-15}$, locate it.

\subsection{Four-Point Inputs}
Beyond $A_2$ we consider $P_{\mu,\rho}=\frac\mu2(\delta_{-\sqrt\rho}+\delta_{\sqrt\rho})+(1-\mu)B_A$. By Lemma~\ref{lem:kkt}, $\partial_\mu I=\Phi(\rho)-\Phi(A^2)$ and $\partial_\rho I=\mu\Phi'(\rho)$ with $\Phi=\Phi_{P_{\mu,\rho}}$, so an optimal input of this form solves $f(\mu,\rho):=(\Phi(\rho)-\Phi(A^2),\Phi'(\rho))=0$. On a parameter cell we enclose a solution in a box $X=x_c+[-h,h]$ by the parametric Krawczyk test $x_c-Yf(x_c,A_m)-Yf_A(A-A_m)+(I-YJ(X))(X-x_c)\subset\operatorname{int}X$, where $Y$ approximates the inverse Jacobian, $J(X)$ encloses the Jacobian over $X$ and the cell, and $f_A$ encloses $\partial_Af(x_c,\cdot)$~\cite{krawczyk1969,moore1977,rump2010}. The KKT conditions are then certified with $G=\Phi-\Phi(A^2)$: $G''<0$ on an anchor $[t_L,t_R]$ containing the $\rho$-range of $X$, cells of type I or V on $[0,t_L]$, D, C, or V on $[t_R,t_k]$, and I or V on $[t_k,A_r^2]$, with type I beyond $A_l^2$ (Lemma~\ref{lem:four} in Appendix~\ref{app:sound}).

\subsection{Near the Third Transition}
\label{sec:cert3}
Close to $A_3$ the KKT function $G=\Phi_{P^*}-\Phi_{P^*}(A^2)$ of the four-point input $P^*=P_{\mu^*,\rho^*}$ nearly vanishes at the origin, and between $0$ and $\rho^*$ it dips only to about $-3.6\times10^{-4}$. Three modifications let the certificates reach $A_3$. First, $G_0(A):=G(0)=\Phi_{P^*}(0)-I(P^*)$ is enclosed along the branch by the mean-value form $G_0(A)\in G_0(A_m)+[G_0'](A-A_m)$, where $G_0(A_m)$ is evaluated on a point Krawczyk box at the midpoint $A_m$ of the cell and $[G_0']$ encloses, over the Krawczyk box of the cell, the total derivative for $x=(\mu,\rho)$:
\begin{equation}
G_0'=\partial_AG_0+\partial_xG_0\cdot x',\qquad x'=-J^{-1}f_A.
\label{eq:G0prime}
\end{equation}
Second, the V-tests use the mean-value form $G(s;\xi)\in G(s;\xi_c)+\nabla_\xi G(s;\Xi)\cdot(\Xi-\xi_c)$ in the parameters $\xi=(\mu,\rho,A)$, where $\Xi$ is the parameter box of the cell and $\xi_c$ its center; this removes most of the overestimation caused by the width of the parameter cell. Third, the cells near the origin follow Lemma~\ref{lem:five} (Appendix~\ref{app:sound}): from the origin, where $G(0)\le0$, forward to the minimum of $G$ on $[0,\rho^*]$ with types D, C, or V, and from there to the anchor with types I or V.

Beyond $A_3$ we consider the five-point family
\begin{equation}
P_{\nu,\mu,\rho}=\nu\delta_0+\tfrac\mu2(\delta_{-\sqrt\rho}+\delta_{\sqrt\rho})+(1-\nu-\mu)B_A .
\label{eq:five}
\end{equation}
By Lemma~\ref{lem:kkt}, $\partial_\nu I=\Phi(0)-\Phi(A^2)$, so an optimal input of this form solves $F:=(f_1,f_2,f_0)=0$ with $f_0=\Phi(0)-\Phi(A^2)$ and $\Phi=\Phi_{P_{\nu,\mu,\rho}}$. For $\nu$ in a cell $[\nu_l,\nu_r]$, which may contain $\nu=0$, a parametric Krawczyk test with $\nu$ as the parameter encloses the solution $(\mu,\rho,A)$ of $F=0$. At $\nu=0$, the Jacobian of $F$ with respect to $(\mu,\rho,A)$ has, by elimination of $(\mu,\rho)$, the determinant $\det J\cdot G_0'(A_3)$, which is nonzero. Derivatives with respect to $\nu$ need tail bounds that do not use a positive lower bound on the weight $\nu$; we bound $1/M$ by $e^{a_j}/w_j$ for one of the other atoms $(w_j,\rho_j)$, $a_j=\rho_j/2$, where $M$ is the factor of the output density in~\eqref{eq:Phi}.

\subsection{Verification of Theorem~\ref{thm:main}}

\begin{proof}[Computer-assisted proof]
(i) is Proposition~\ref{prop:binary} with opposite certified signs of $\cH_0$ (negative on the left, positive on the right) at the rational endpoints of a bracket of width $6.1\times10^{-9}$; rounded outward, this bracket is the stated interval. (ii) On $[1.6659,1.73]$, $321$ cells ($320$ of width $2\times10^{-4}$ and one of width $10^{-4}$) certify the upper inequality $\cH_0(A)<\cK\lam_0/2$ in Lemma~\ref{lem:gate}; the lower inequality $\cH_0(A)>0$ holds for $A>A_1$ by Proposition~\ref{prop:binary}, so the condition holds on $(A_1,1.73]$; Algorithm~1 covers $[1.73,2.895]$ with $313$ cells and $30{,}732$ $s$-cells; and Algorithm~3 covers $[2.895,2.9074]$ with $13$ cells and $7{,}580$ $s$-cells, on which $g<0$ is certified on $100$ sub-cells. On the gap cell $[2.9074,2.9075]$, Algorithm~3 ($146$ $s$-cells) gives optimality wherever $g\le0$, and by (iv) this is $[2.9074,A_2]$. (iv) The first cells of Algorithms~1 and~3 contain $s=0$ with $g<0$ certified, and on the range of the analytic condition $\Phi_{\lam^*}$ is strictly convex with equal end values, so $g<0$ on $(A_1,2.9074]$. On the gap cell the enclosure of~\eqref{eq:gprime} is $[0.16601,0.17182]$. Algorithm~2 certifies $g>0$ on $247$ cells covering $[2.9075,3.2]$. Twenty steps of bisection with certified signs of $g$ give a bracket of $A_2$ with rational end points and width $9.5\times10^{-11}$, at which $g\in[-1.0383,-1.0364]\times10^{-12}$ and $g\in[1.5070,1.5073]\times10^{-11}$; rounded outward, it is the stated interval. (iii) An optimal input is symmetric and contains $\pm A$~\cite{smith1971}; with at most three points it is $B_A$, excluded by (i), or some $P_\lam$, excluded by (iv) and Proposition~\ref{prop:exit} for $A\le3.2$; for $A\in(A_2,3.98]$, (iii) also follows from (v), since the CAID is unique. (v) The Krawczyk test and the KKT cells succeed on $4{,}393$ cells covering $[2.907464074,3.98]$, with $1{,}681{,}495$ $s$-cells and cell widths that shrink geometrically towards $A_2$ and, beyond $3.9$, towards $A_3$. The first box, at $A=2.907464074$, is $\mu\in[0.276500986,0.276500987]$, $\rho\in[3.19\times10^{-9},6.06\times10^{-9}]$. The remaining interval $(A_2,2.907464074)$ is covered by parametrizing the branch by $\rho$ instead of $A$: for $\rho\in[0,\rho_1]$, $\rho_1=6\times10^{-9}$, a parametric Krawczyk test solves $f(\mu,\rho,A)=0$ for $(\mu,A)$. At the splitting point $(\lam_2,0,A_2)$, the Jacobian determinant with respect to $(\mu,A)$ is $J_{11}\partial_Af_2-J_{21}\partial_Af_1=J_{11}g'(A_2)\ne0$ (Section~\ref{sec:laws}). By continuity the determinant remains nonzero in a neighborhood of that point, so the parametrization includes $\rho=0$. On both $\rho$-cells the KKT tests of Lemma~\ref{lem:four} succeed with anchor $[0,0.05]$ ($288$ $s$-cells). The first $A$-box lies inside the gap cell. At $\rho=0$, $f_1=0$ gives the unique ternary weight $\lam^*(A)$; then $f_2=0$ gives $g(A)=0$, so $A(0)=A_2$ by (iv). A point box at the shared edge lies inside both cell boxes, so uniqueness identifies their solutions and $A(\rho)$ is continuous. A point box inside the last cell box gives $A(\rho_1)\in[2.9074640753048,2.9074640753069]$; hence every $A\in(A_2,A(\rho_1)]$ is $A(\rho)$ for some $\rho\in(0,\rho_1]$, and the corresponding four-point input is the CAID.
\end{proof}

\begin{figure*}[t]
\centering
\includegraphics[width=\textwidth]{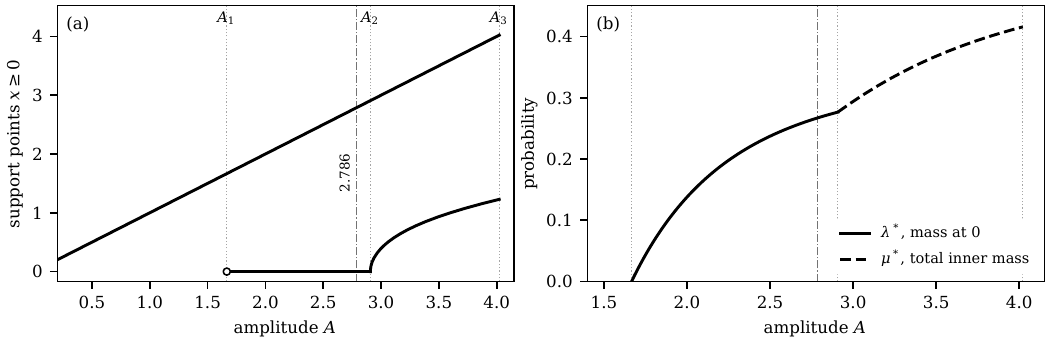}
\caption{Scalar optimal input through $A_3$: (a) nonnegative support points and (b) mass at zero or total mass of the inner pair. Theorems~\ref{thm:main} and~\ref{thm:third} certify the support regimes; the plotted coordinates are numerical. The open circle in (a) excludes the origin at $A_1$. Vertical dotted lines mark the transitions; the dash-dotted line marks the estimate $2.786$ in~\cite{sharma2010}. The five-point regime certified on $(A_3,A_3^+]$ is not plotted.}
\label{fig:support}
\end{figure*}

\begin{figure*}[t]
\centering
\includegraphics[width=\textwidth]{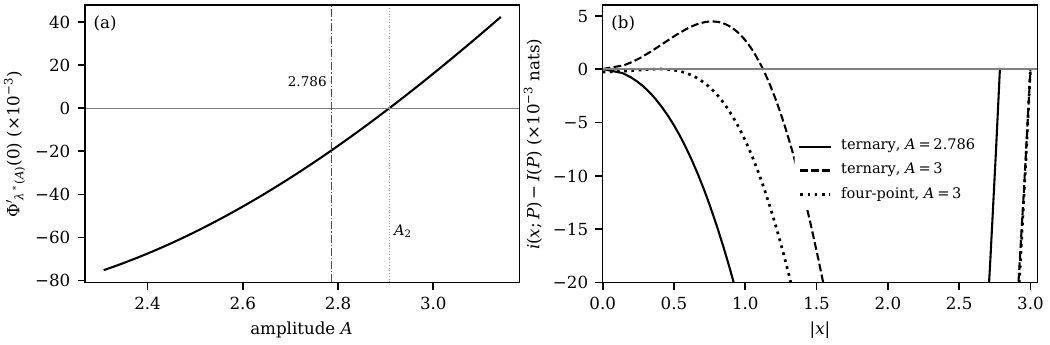}
\caption{(a) $g(A)=\Phi'_{\lam^*(A)}(0)$, whose zero is $A_2$; at $2.786$ it is negative. (b) KKT functions $i(x;P)-I(P)$: the ternary input at $A=2.786$ (nonpositive, optimal), the ternary input at $A=3$ (positive near $|x|\approx0.8$, not optimal), and the four-point input at $A=3$ (nonpositive, zero at $|x|=r^*$ and $|x|=3$). The dash-dotted line in (a) marks the estimate in~\cite{sharma2010}. All curves are numerical.}
\label{fig:kkt}
\end{figure*}

\section{Transition Laws}
\label{sec:laws}

Let $\delta$ denote the distance to a transition, $C_2(A)=I(B_A)$, and $C_3(A)=\max_{0\le\lam<1}I(P_\lam)$ the largest rate of an input supported on $\{0,\pm A\}$. For $A>A_1$, $C_3(A)=I(P_{\lam^*(A)})$, since $\lam\mapsto I(P_\lam)$ is concave with derivative $\cH(\lam,A)$. For the four-point family let $f(\mu,\rho,A)=(f_1,f_2)=(\Phi(\rho)-\Phi(A^2),\Phi'(\rho))$ with $\Phi=\Phi_{P_{\mu,\rho}}$, as in Section~\ref{sec:cert}. Since $e^{-\rho/2}\cosh(\sqrt{\rho q})$ is an entire function of $\rho$, the output density of $P_{\mu,\rho}$ extends analytically to small negative $\rho$; for $|\rho|$ small it stays positive, and $f$ is real-analytic in $(\mu,\rho,A)$ near $\rho=0$ (differentiate under the expectation). At $\rho=0$, $P_{\mu,0}=P_\mu$, so $f(\mu,0,A)=(\cH(\mu,A),\Phi'_\mu(0))$. Write $J=\partial(f_1,f_2)/\partial(\mu,\rho)$ and $f_A=\partial f/\partial A$.

\begin{proposition}[Onset of mass at the origin]\label{prop:appear}
Let $\cH_0(A_1)=0$, $\cH_0'(A_1)>0$, and $\cK_0:=-\partial_\lam\cH(0,A_1)>0$, and let $\delta=A-A_1$. Then, as $\delta\downarrow0$,
\begin{gather*}
\lam^*(A)=\ell_1\delta+O(\delta^2),\\ C_3(A)-C_2(A)=K_1\delta^2+O(\delta^3),\\
\ell_1=\frac{\cH_0'(A_1)}{\cK_0},\qquad K_1=\frac{\cH_0'(A_1)^2}{2\cK_0}.
\end{gather*}
\end{proposition}
\begin{proof}
The function $\cH$ is real-analytic near $(0,A_1)$, including small negative $\lam$, and $\partial_\lam\cH(0,A_1)=-\cK_0\neq0$. Its analytic stationary root $\bar\lam(A)$ has $\bar\lam'(A_1)=\cH_0'(A_1)/\cK_0>0$. On the right $\bar\lam=\lam^*$; on the left $\bar\lam<0$ is not an admissible mass. Define the analytic continuation of the stationary value by
\[
\bar C_3(A)=C_2(A)+\int_0^{\bar\lam(A)}\cH(t,A)\,dt .
\]
By Lemma~\ref{lem:kkt}, $C_3=\bar C_3$ just to the right of $A_1$, whereas $C_3=C_2$ just to the left by optimality of $B_A$. On the right, $\cH(\lam,A)=\cH_0(A)-\cK_0\lam+O(\lam^2+|\delta|\lam)$ and $\lam^*=O(\delta)$. Integration gives $C_3-C_2=\cH_0(A)^2/(2\cK_0)+O(\delta^3)$, with $\cH_0(A)=\cH_0'(A_1)\delta+O(\delta^2)$.
\end{proof}

\begin{proposition}[Splitting of the mass at the origin]\label{prop:split}
Let $\lam_2=\lam^*(A_2)$, so that $f(\lam_2,0,A_2)=0$, and evaluate $J$ and $f_A$ at $(\lam_2,0,A_2)$. Suppose that
\begin{gather*}
g'(A_2)=\partial_Af_2-\frac{J_{21}\,\partial_Af_1}{J_{11}}>0,\\
D:=-\frac{\det J}{J_{11}}=-J_{22}+\frac{J_{12}J_{21}}{J_{11}}>0 .
\end{gather*}
Then there is a real-analytic map $A\mapsto(\tilde\mu(A),\tilde\rho(A))$ near $A_2$ with $f(\tilde\mu,\tilde\rho,A)=0$, $\tilde\mu(A_2)=\lam_2$, and, on the admissible side $\delta=A-A_2\downarrow0$,
\begin{gather*}
\tilde\rho(A)=c\,\delta+O(\delta^2),\\
I(P_{\tilde\mu,\tilde\rho})-C_3(A)=K_2\delta^2+O(\delta^3),\\
c=\frac{g'(A_2)}{D},\qquad K_2=\frac{\lam_2\,g'(A_2)\,c}{2}=\frac{\lam_2\,g'(A_2)^2}{2D}.
\end{gather*}
\end{proposition}
\begin{proof}
$J_{11}=\partial_\mu\cH=-\cK_{\lam_2}<0$ and $\det J=-J_{11}D\neq0$, so the implicit function theorem gives the analytic branch, and $(\tilde\mu',\tilde\rho')=-J^{-1}f_A$; its second component is $(J_{21}\partial_Af_1-J_{11}\partial_Af_2)/\det J=g'(A_2)/D$, where $g'(A_2)$ is the derivative of $g(A)=f_2(\lam^*(A),0,A)$ because $\lam^{*\prime}=-\partial_Af_1/J_{11}$. For the rate, fix $A$ and let $\hat\mu(\rho)$ solve $f_1(\hat\mu,\rho,A)=0$ near $\lam^*(A)$; since $\mu\mapsto I(P_{\mu,\rho})$ is concave with derivative $f_1$, $\hat\mu(\rho)$ maximizes it for small $\rho\ge0$, and $W(\rho)=I(P_{\hat\mu(\rho),\rho})$ satisfies $W(0)=C_3(A)$ and $\partial_\mu I=f_1=0$ at $\mu=\hat\mu(\rho)$. Hence
\[
W'(\rho)=\left.\partial_\rho I\right|_{\mu=\hat\mu(\rho)}
          =\hat\mu(\rho) f_2(\hat\mu(\rho),\rho,A)
\]
by Section~\ref{sec:cert}. The function $\varphi(\rho)=f_2(\hat\mu(\rho),\rho,A)$ satisfies $\varphi(0)=g(A)=g'(A_2)\delta+O(\delta^2)$ and $\varphi'(0)=J_{22}-J_{21}J_{12}/J_{11}+O(\delta)=-D+O(\delta)$, and it vanishes at $\tilde\rho(A)$. Hence $W(\tilde\rho)-W(0)=\int_0^{\tilde\rho}\hat\mu\varphi\,d\rho=\lam_2\big(g\tilde\rho-D\tilde\rho^2/2\big)+O(\delta^3)=\lam_2g^2/(2D)+O(\delta^3)$.
\end{proof}

The rate formulas have a simple reading. Near $A_1$, the derivative of the rate in the direction of the new atom is $\cH_0\approx\cH_0'\delta$ and the curvature in that direction is $-\cK_0$; near $A_2$, the derivative of the rate in the direction of moving the atom at the origin outwards is $\lam_2\,g\approx\lam_2g'\delta$, and the effective curvature in that direction is $-\lam_2D$. In both cases the leading gain is half the squared slope divided by the magnitude of the curvature.

\begingroup\sloppy
\begin{theorem}\label{thm:laws}
For the scalar amplitude-constrained Gaussian channel:
\begin{enumerate}
\item[(i)] The hypotheses of Proposition~\ref{prop:appear} hold, with $\ell_1\in[0.5620292,0.5620294]$ and $K_1^{\rm bit}\in[0.2931159,0.2931161]$ bit; as $A\downarrow A_1$, $C(A)-C_2(A)=K_1(A-A_1)^2+O((A-A_1)^3)$.
\item[(ii)] The hypotheses of Proposition~\ref{prop:split} hold, with $g'(A_2)\in[0.1689168,0.1689169]$, $D\in[0.0972233,0.0972235]$, $c\in[1.7374087,1.7374097]$, and $K_2^{\rm bit}\in[0.05853507,0.05853512]$ bit. For $A$ in a right neighborhood of $A_2$, the CAID of Theorem~\ref{thm:main}(v) is the four-point input $P_{\tilde\mu(A),\tilde\rho(A)}$ of Proposition~\ref{prop:split}; its inner radius satisfies $r^*=\sqrt{c\,\delta}\,(1+O(\delta))$, its inner mass $\tilde\mu(A)=\lam_2+O(\delta)$, and $C(A)-C_3(A)=K_2\delta^2+O(\delta^3)$, $\delta=A-A_2$.
\item[(iii)] $C$ is continuously differentiable near $A_1$ and $A_2$, its one-sided second derivatives exist there, and $C''(A_i+)-C''(A_i-)=2K_i>0$, $i=1,2$. In particular, $C$ is not twice differentiable at $A_1$ and $A_2$.
\end{enumerate}
\end{theorem}
\endgroup
\begin{proof}[Computer-assisted proof]
(i) We enclose $\cH_0'(A)$ and $\cK_0=\E_{u\sim\mathrm{Gamma}(1/2,1)}[(1-p)^2/p]$, $p=e^{-a}\mathsf F_0(au)$, over the certified bracket of $A_1$, with the tail bounds of Appendix~\ref{app:formulas}; both are positive. For $A\in(A_1,A_2]$, $C=C_3$ by Theorem~\ref{thm:main}(ii).

(ii) Over the box $A\in[A_2^-,A_2^+]$ (the rational bracket of $A_2$), $\lam$ in a certified bracket of $\lam^*$ on this box (width $4\times10^{-9}$), and $\rho=0$, we enclose $J$ and $f_A$ with the routines of the Krawczyk test (Section~\ref{sec:cert}); this gives the stated intervals and, in particular, $g'(A_2)>0$ and $D>0$. By Proposition~\ref{prop:split}, the branch exists and $\tilde\rho(A)>0$ for small $\delta>0$. It remains to identify the CAID with this branch. By Theorem~\ref{thm:main}(v), for $A\in(A_2,3.98]$ the CAID is a four-point input $P_{\mu^*(A),\rho^*(A)}$, and near $A_2$ its parameters lie in the Krawczyk boxes of the $\rho$-parametrized certificate. They converge to $(\lam_2,0)$ as $A\downarrow A_2$: a limit point $(\bar\mu,\bar\rho)$ with $\bar\rho>0$ would, by the same certificate, give a four-point CAID at $A_2$, contradicting Theorem~\ref{thm:main}(ii) and uniqueness. Since $f$ has a unique zero near $(\lam_2,0,A_2)$ for each $A$ (implicit function theorem), $(\mu^*,\rho^*)=(\tilde\mu,\tilde\rho)$ for $A$ close to $A_2$. The remaining claims follow from Proposition~\ref{prop:split}.

(iii) At $A_1$ the left and right pieces of $C$ have analytic extensions $C_2$ and $\bar C_3$, respectively; the constrained value $C_3$ itself is not analytic across $A_1$. At $A_2$ the ternary mass is interior, so $C_3$ is analytic there; the right piece extends along the stationary splitting branch of Proposition~\ref{prop:split}. By (i) and (ii), the right and left analytic extensions differ by $K_i\delta^2+O(\delta^3)$. Their values and first derivatives agree at $A_i$, and their second derivatives differ by $2K_i$.
\end{proof}

The enclosures of Theorem~\ref{thm:laws} take less than a second; the finite-increment floating-point estimates differ from $\ell_1$ and $c$ by less than $10^{-4}$ ($\lam^*/\delta=0.56197$ and $\tilde\rho/\delta=1.73735$ at $\delta=10^{-4}$), and the certified value of $g'(A_2)$ refines the enclosure $[0.1660,0.1719]$ of Theorem~\ref{thm:main}(iv), which holds on the whole gap cell. The inner mass moves at rate $\tilde\mu'(A_2)\in[0.196314,0.196315]$. By Theorem~\ref{thm:laws}(ii), as $A\downarrow A_2$ the ratio $\rho^*(A)/(A-A_2)$ tends to $c\approx1.73741$, and $r^*/\sqrt{A-A_2}$ tends to $\sqrt c\approx1.31811$.

\begin{figure*}[t]
\centering
\includegraphics[width=0.95\textwidth]{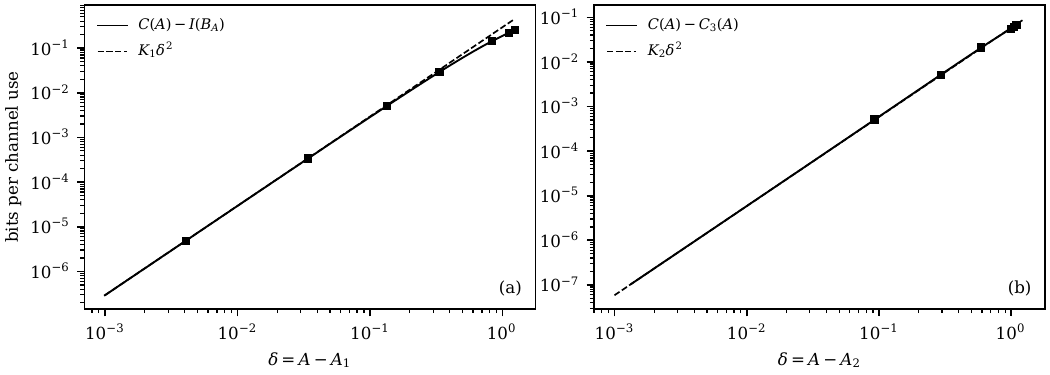}
\caption{Rate loss in bits per channel use versus distance to the scalar transition: (a) $C(A)-C_2(A)$ above $A_1$ and (b) $C(A)-C_3(A)$ above $A_2$. Solid curves are numerical; dashed curves show $K_i^{\rm bit}\delta^2$ using midpoints of the coefficient enclosures in Theorem~\ref{thm:laws}. Squares mark midpoints of certified loss intervals: seven in (a) and six in (b). Table~\ref{tab:loss} lists ten of these intervals.}
\label{fig:laws}
\end{figure*}

Table~\ref{tab:loss} compares the certified losses with their quadratic predictions at selected amplitudes; Fig.~\ref{fig:laws} shows how the losses vary with distance to the transition.

\begin{table}[t]
\centering
\caption{Certified rate lost by retaining the previous constellation family, in bits, compared with the transition laws. Upper part: $C(A)-C_2(A)$ in the ternary regime; lower part: $C(A)-C_3(A)$ at the certified four-point inputs. The last column is $K_i^{\rm bit}\delta^2$, $\delta=A-A_i$.}
\label{tab:loss}
\begin{tabular}{cccc}
\toprule
$A$ & $\delta$ & certified loss & $K_i^{\rm bit}\delta^2$\\
\midrule
$1.7$ & $0.0341$ & $[3.368,3.369]\times10^{-4}$ & $3.403\times10^{-4}$\\
$1.8$ & $0.1341$ & $[5.049,5.050]\times10^{-3}$ & $5.269\times10^{-3}$\\
$2.0$ & $0.3341$ & $[0.02906,0.02907]$ & $0.0327$\\
$2.5$ & $0.8341$ & $[0.1429,0.1430]$ & $0.2039$\\
$2.9074$ & $1.2415$ & $[0.2526,0.2527]$ & $0.4518$\\
\midrule
$3.0$ & $0.0925$ & $[4.863,5.300]\times10^{-4}$ & $5.012\times10^{-4}$\\
$3.2$ & $0.2925$ & $[5.126,5.174]\times10^{-3}$ & $5.009\times10^{-3}$\\
$3.5$ & $0.5925$ & $[0.02091,0.02098]$ & $0.0206$\\
$3.9$ & $0.9925$ & $[0.05575,0.05582]$ & $0.0577$\\
$4.0$ & $1.0925$ & $[0.06632,0.06639]$ & $0.0699$\\
\bottomrule
\end{tabular}
\end{table}

Let $L$ denote the actual rate loss and $\widehat L=K_i^{\rm bit}\delta^2$ its quadratic prediction, both in bits. For each of the five four-point inputs in Table~\ref{tab:loss} ($A=3.0$ through $4.0$), the certified enclosure of $L$ gives $|\widehat L-L|/L<6\%$. For the onset of mass at $A_1$, this error is below $5\%$ at $A=1.7$ and $1.8$ and below $13\%$ at $A=2.0$; the prediction increasingly overestimates the loss at the two larger tabulated amplitudes. These comparisons are pointwise; the continuous curves in Fig.~\ref{fig:laws} are numerical. The quadratic laws predict a loss of $10^{-3}$ bit at about $A_1+0.058$ for the binary input and $A_2+0.13$ for the ternary input. Thus a small delay in switching constellations need not entail a large rate penalty. At the third transition, mass returns to the origin while the existing points readjust (Section~\ref{sec:third}).

\section{The Third Transition}
\label{sec:third}

Let $x^*(A)=(\mu^*(A),\rho^*(A))$ be the four-point branch of Theorem~\ref{thm:main}(v), $P^*=P_{\mu^*,\rho^*}$, $C_4(A)=I(P^*)$, and $G_0(A)=\Phi_{P^*}(0)-I(P^*)$ as in Section~\ref{sec:cert3}. In the five-point family~\eqref{eq:five}, the output density stays positive for small negative $\nu$, and $F=(f_1,f_2,f_0)$ is real-analytic in $(\nu,\mu,\rho,A)$ near $\nu=0$; at $\nu=0$, $f_0=G_0$ on the branch.

\begin{proposition}[Re-entry of mass at the origin]\label{prop:reentry}
Let $G_0(A_3)=0$, let $J$ be nonsingular at $(x^*(A_3),A_3)$, and suppose that, at $(0,x^*(A_3),A_3)$,
\begin{gather*}
h_3:=G_0'(A_3)>0,\\
\cK_{\mathrm{eff}}:=-\big(\partial_\nu f_0-\partial_xf_0\,J^{-1}\partial_\nu f\big)>0,
\end{gather*}
where $f=(f_1,f_2)$ and $x=(\mu,\rho)$. Then there is a real-analytic map $A\mapsto(\tilde\nu(A),\tilde x(A))$ near $A_3$ with $F(\tilde\nu,\tilde x,A)=0$ and $\tilde\nu(A_3)=0$, and, on the admissible side $\delta=A-A_3\downarrow0$,
\begin{gather*}
\tilde\nu(A)=\ell_3\delta+O(\delta^2),\\ I(P_{\tilde\nu,\tilde x})-C_4(A)=K_3\delta^2+O(\delta^3),\\
\ell_3=\frac{h_3}{\cK_{\mathrm{eff}}},\qquad K_3=\frac{h_3^2}{2\cK_{\mathrm{eff}}}.
\end{gather*}
\end{proposition}
\begin{proof}
Since $J$ is nonsingular, the implicit function theorem gives a solution $\hat x(\nu,A)$ of $f(\nu,\hat x,A)=0$ near $(0,x^*(A_3),A_3)$, analytic in $(\nu,A)$, with $\hat x(0,A)=x^*(A)$ and $\partial_\nu\hat x=-J^{-1}\partial_\nu f$. Fix $A$ and let $W(\nu)=I(P_{\nu,\hat x(\nu,A)})$. By Lemma~\ref{lem:kkt}, $\partial_\mu I=f_1$, $\partial_\rho I=\mu f_2$, and $\partial_\nu I=f_0$, and $f=0$ along $\hat x$, so $W'(\nu)=\varphi(\nu):=f_0(\nu,\hat x(\nu,A),A)$. Here $W(0)=C_4(A)$, $\varphi(0)=G_0(A)=h_3\delta+O(\delta^2)$, and $\varphi'(0)=\partial_\nu f_0+\partial_xf_0\,\partial_\nu\hat x=-\cK_{\mathrm{eff}}+O(\delta)$. Since $\varphi$ is analytic in $(\nu,A)$ with $\partial_\nu\varphi\neq0$ at $(0,A_3)$, its zero $\tilde\nu(A)$ is analytic, with $\tilde\nu=h_3\delta/\cK_{\mathrm{eff}}+O(\delta^2)$; let $\tilde x(A)=\hat x(\tilde\nu(A),A)$. Finally, $W(\tilde\nu)-W(0)=\int_0^{\tilde\nu}\varphi\,d\nu=G_0\tilde\nu-\cK_{\mathrm{eff}}\tilde\nu^2/2+O(\delta^3)=h_3^2\delta^2/(2\cK_{\mathrm{eff}})+O(\delta^3)$.
\end{proof}

Proposition~\ref{prop:appear} is the special case without the inner pair. The effective curvature $\cK_{\mathrm{eff}}$ is a Schur complement: it is the curvature of the rate in the direction of the new mass after the other parameters have been re-optimized. The reading of Section~\ref{sec:laws} persists: the gain is half the squared slope divided by the effective curvature.

\begingroup\sloppy
\begin{theorem}\label{thm:third}
The third scalar transition point satisfies $A_3\in[4.0212753203,4.0212753243]$. Let $A_3^+=4.0212754757$.
\begin{enumerate}
\item[(i)] For every $A\in(A_2,A_3]$, the CAID is the four-point input $P^*$ of Theorem~\ref{thm:main}(v). Moreover, $\Phi_{P^*}(0)<C(A)$ for $A<A_3$, and $\Phi_{P^*}(0)=C(A_3)$ at $A_3$.
\item[(ii)] On $[4.021275,4.021276]$, $G_0$ is strictly increasing with $G_0'\in[0.03398,0.03460]$, and $A_3$ is its unique zero there; for $A$ in this interval, $P^*$ is capacity-achieving if and only if $A\le A_3$.
\item[(iii)] For every $A\in(A_3,A_3^+]$, the CAID is a five-point input $P_{\nu,\mu,\rho}$ with $\nu>0$: a mass point re-enters at the origin.
\item[(iv)] The hypotheses of Proposition~\ref{prop:reentry} hold, with
\begin{gather*}
h_3\in[0.0342926,0.0342952],\\ \cK_{\mathrm{eff}}\in[0.0526643,0.0526778],\\
\ell_3\in[0.65098,0.65121],\\ K_3^{\rm bit}\in[0.0161034,0.0161099]\text{ bit}.
\end{gather*}
As $A\downarrow A_3$, the mass at the origin is $\ell_3\delta+O(\delta^2)$ and $C(A)-C_4(A)=K_3\delta^2+O(\delta^3)$. Hence $C$ is continuously differentiable near $A_3$, its one-sided second derivatives exist, and $C''(A_3+)-C''(A_3-)=2K_3>0$.
\end{enumerate}
\end{theorem}
\endgroup
\begin{proof}[Computer-assisted proof]
(i) On $(A_2,3.98]$, the CAID is $P^*$ by Theorem~\ref{thm:main}(v), and the cells used there give $G_0<0$: near the origin they are of type I (strictly increasing $G$) or V, and near $A_2$ the strictly concave anchor $[0,0.05]$ contains $\rho^*>0$, where $G$ attains its maximum $0$. On $[3.98,4.021275]$, the Krawczyk test, the enclosure $G_0<0$ of Section~\ref{sec:cert3}, and the cells of Lemma~\ref{lem:five} succeed on $917$ cells with $1{,}120{,}969$ $s$-cells. Most cells have width $5\times10^{-5}$ ($799$ cells) or $10^{-5}$ ($93$ cells). The remaining cells are shorter. They occur at the ends of the parallel blocks or, within $1.7\times10^{-4}$ of $A_3$, have widths of about $0.3(A_3-A)$, down to $9.4\times10^{-8}$. The remaining claims follow from (ii).
(ii) On the gap cell $[4.021275,4.021276]$, the Krawczyk test and the cells of Lemma~\ref{lem:five} succeed except for the sign of $G_0$ ($1{,}629$ $s$-cells), and~\eqref{eq:G0prime} gives the stated enclosure of $G_0'$ over the cell. Hence $G_0$ has at most one zero on the cell, $P^*$ is optimal where $G_0\le0$ by Lemma~\ref{lem:five}, and where $G_0>0$ it violates the KKT condition of Lemma~\ref{lem:kkt} at the origin. In each of the first eight bisection steps, a point Krawczyk box certifies the sign of $G_0$ at the midpoint, and the bracket is halved. The ninth midpoint enclosure contains zero, so we retain the bracket of width $3.9\times10^{-9}$. At its rational left and right end points, $G_0\in[-1.114\times10^{-10},-1.828\times10^{-11}]$ and $G_0\in[2.258\times10^{-11},1.157\times10^{-10}]$, respectively. Rounding the end points outward gives the stated interval.
(iii) For $\nu\in[0,10^{-7}]$, divided into four cells, the parametric Krawczyk test of Section~\ref{sec:cert3} and the cells of Lemma~\ref{lem:five}, now with $G(0)=f_0=0$, succeed ($1{,}625$ $s$-cells per cell). The box of the first cell lies, in $(\mu,\rho)$, inside the Krawczyk box of the gap cell and, in $A$, inside the gap cell. At $\nu=0$ its solution is therefore the four-point branch at a zero of $G_0$, i.e., $A(0)=A_3$ by (ii). Point tests at the three inner edges show that the solutions of adjacent cells coincide there, so $\nu\mapsto A(\nu)$ is continuous on $[0,10^{-7}]$, and a point box inside the last cell box gives $A(10^{-7})\in[4.02127547579408,4.02127547579610]$. Hence every $A\in(A_3,A_3^+]$ equals $A(\nu)$ for some $\nu\in(0,10^{-7}]$, and the five-point input with these parameters satisfies the KKT conditions, so it is the CAID.
(iv) Over the bracket of $A_3$ and a Krawczyk box of $x^*$, we enclose $h_3$ by~\eqref{eq:G0prime} and $\cK_{\mathrm{eff}}$ with the tail bounds of Section~\ref{sec:cert3}; both are positive. As $A\downarrow A_3$, the branch $(\tilde\nu,\tilde x,A)$ of Proposition~\ref{prop:reentry} tends to $(0,x^*(A_3),A_3)$, which is the solution at $\nu=0$ of the first cell of (iii) and lies in the interior of its box. By uniqueness in that box, the branch is the CAID of (iii) for $A$ close to $A_3$. Thus $C=C_4$ on the left of $A_3$ and $C=I(P_{\tilde\nu,\tilde x})$ on the right, both real-analytic, and the claims follow from Proposition~\ref{prop:reentry} as in the proof of Theorem~\ref{thm:laws}(iii).
\end{proof}

At $A_3$, the CAID is $\{\pm r^*,\pm A\}$ with $r^*\approx1.22896$ and inner mass $\mu^*\approx0.41570$; after the re-entry, the new mass at the origin grows at rate $\ell_3\approx0.651$. The constants of the rate loss decrease from transition to transition, $K_1^{\rm bit}\approx0.293$, $K_2^{\rm bit}\approx0.0585$, and $K_3^{\rm bit}\approx0.0161$ bit, so that, in the sense of Section~\ref{sec:laws}, the transitions become less critical as $A$ grows. In peak signal-to-noise ratio $A^2$, the three transitions lie at $4.43$, $9.27$, and $12.09$~dB.

\section{The Quadrature Channel}
\label{sec:quad}

Consider $Y=X+Z$ with $X,Y\in\mathbb R^2$ (complex baseband), $Z\sim\mathcal N(0,I_2)$, and $\|X\|\le A$. The CAID is unique and invariant under rotations, and its amplitude is discrete~\cite{shamai1995,rassouli2016}; it is a mixture of uniform distributions $U_r$ on circles of radius $r\le A$. For such inputs $i(x;P)$ depends on $s=\|x\|^2$, and Sections~\ref{sec:kkt}--\ref{sec:laws} carry over with the following changes: $Q_{m,s}$ is replaced by $Q_{m+1,s}$ and $\cosh(A\sqrt q)$ by $I_0(A\sqrt q)={}_0F_1(;1;A^2q/4)$; the binary input $B_A$ becomes the constant-envelope input $U_A$, the ternary input becomes the constant envelope with a zero symbol, $P_\lam=\lam\delta_0+(1-\lam)U_A$, and the four-point input becomes the two-ring input $P_{\mu,\rho}=\mu U_{\sqrt\rho}+(1-\mu)U_A$. Lemmas~\ref{lem:kkt}, \ref{lem:sound}, and~\ref{lem:four} and Propositions~\ref{prop:exit}--\ref{prop:split} hold verbatim, and the formulas of Appendix~\ref{app:formulas} hold with $\eta=1$. Lemma~\ref{lem:gate} applies directly to $I_0$. The curvature estimate in its proof also gives the endpoint criterion: $U_A$ is optimal exactly when $\cH_0(A)\le0$, since the common value on the outer circle is $I(U_A)$ and $\Phi_0$ is strictly convex in $s$. The unique constant-envelope threshold is the one characterized in~\cite[Thms.~2 and~3]{dytso2019}; on its right $\cH_0(A)>0$.

\begingroup\sloppy
\begin{theorem}\label{thm:quad}
The first two quadrature transition points satisfy $A_1\in[2.4533899536,2.4533899598]$ and $A_2\in[3.420348444271,3.420348444367]$.
\begin{enumerate}
\item[(i)] The constant-envelope input $U_A$ is capacity-achieving if and only if $A\le A_1$.
\item[(ii)] For every $A\in(A_1,A_2]$, the CAID is $P_{\lam^*(A)}$, with $0<\lam^*(A)<1$.
\item[(iii)] The function $g(A)=\Phi'_{\lam^*(A)}(0)$ is negative on $(A_1,3.4203]$, strictly increasing on $[3.4203,3.4204]$ with $g'\in[0.1915,0.1994]$, and positive on $[3.4204,3.8]$; hence no input $P_\lam$ is capacity-achieving for $A\in(A_2,3.8]$.
\item[(iv)] For every $A\in(A_2,4]$, the CAID is a two-ring input $P_{\mu^*,\rho^*}$ with $0<\rho^*<A^2$.
\item[(v)] The hypotheses of Propositions~\ref{prop:appear} and~\ref{prop:split} hold, with $\ell_1\in[0.1857645,0.1857647]$, $K_1^{\rm bit}\in[0.1307337,0.1307339]$ bit, $g'(A_2)\in[0.1954794,0.1954796]$, $D\in[0.0624378,0.0624379]$, $c\in[3.130785,3.130788]$, and $K_2^{\rm bit}\in[0.0406259,0.0406261]$ bit, and the conclusions of Theorem~\ref{thm:laws} hold with these constants. In particular, $C$ is continuously differentiable but not twice differentiable at $A_1$ and $A_2$.
\end{enumerate}
\end{theorem}
\endgroup
\begin{proof}[Computer-assisted proof]
The certificates are those of Theorems~\ref{thm:main} and~\ref{thm:laws} with $n=2$. (i) Certified signs of $\cH_0$ at the end points of a bracket of width $6.1\times10^{-9}$. (ii) The $184$ cells covering $[2.4533,2.49]$ certify the upper inequality of Lemma~\ref{lem:gate}; its lower inequality holds for $A>A_1$. The remaining certificates are Algorithm~1 on $[2.49,3.41]$ ($249$ cells, $26{,}865$ $s$-cells), Algorithm~3 on $[3.41,3.4203]$ ($11$ cells, $7{,}650$ $s$-cells), and the gap cell $[3.4203,3.4204]$ ($140$ $s$-cells). (iii) Algorithm~2 on $[3.4204,3.8]$ ($279$ cells), the transversality certificate on the gap cell, and twenty steps of certified bisection, which give a bracket of $A_2$ of width $9.5\times10^{-11}$. (iv) The $\rho$-parametrized certificate on $\rho\in[0,6\times10^{-4}]$ ($16$ cells, linked by point tests at the inner edges, with the endpoint box inside the last cell box) covers $(A_2,A(\rho_1)]$ with $A(\rho_1)\in[3.4205400985,3.4205400986]$, and $1{,}449$ cells with $271{,}275$ $s$-cells cover $[3.4205,4]$. (v) As in the proof of Theorem~\ref{thm:laws}, with $\lam^*(A_2)\in[0.0920247,0.0920248]$.
\end{proof}

Table~\ref{tab:const} summarizes the locations and leading coefficients for all five transitions.

\begin{table*}[t]
\centering
\caption{Certified transition locations and leading coefficients. The onset coefficient is $\ell$ for mass at the origin and $c$ for squared inner radius. The leading rate loss in bits is $K^{\rm bit}\delta^2$. Coefficients are rounded; Theorems~\ref{thm:laws}--\ref{thm:quad} give the certified enclosures.}
\label{tab:const}
\small
\setlength{\tabcolsep}{4pt}
\begin{tabular}{llccc}
\toprule
channel & transition & location & onset coefficient & $K^{\rm bit}$\\
\midrule
scalar & appearance at $0$ & $[1.6659256408,1.6659256470]$ & $\ell_1=0.5620293$ & $0.2931160$\\
scalar & splitting of $0$ & $[2.907464071846,2.907464071942]$ & $c=1.737409$ & $0.0585351$\\
scalar & re-entry at $0$ & $[4.0212753203,4.0212753243]$ & $\ell_3=0.6511$ & $0.01611$\\
quadrature & appearance at $0$ & $[2.4533899536,2.4533899598]$ & $\ell_1=0.1857646$ & $0.1307338$\\
quadrature & splitting of $0$ & $[3.420348444271,3.420348444367]$ & $c=3.13079$ & $0.0406260$\\
\bottomrule
\end{tabular}
\end{table*}

\emph{Comparison with earlier numerical thresholds.} Theorem~\ref{thm:quad}(i) is the case $n=2$ of~\cite[Thm.~2]{dytso2019}. Our bracket refines the numerical value $\bar R_2\approx2.454$ in~\cite[Table~I]{dytso2019}. Shamai and Bar-David~\cite[Sec.~III-B and Table~I]{shamai1995} use unit noise variance per real component and the peak-SNR parameter $\rho_p=A^2$. In their units our bracket gives $A_1^2\in[6.019122,6.019123]$, compared with their numerical threshold $6.022$; both give $7.8$~dB to one decimal place.

The capacity column of~\cite[Table~I]{shamai1995} assigns three radii of positive probability to approximately $9.4$--$10.7$~dB, counting the origin when it has positive mass. Theorem~\ref{thm:quad}(ii) and (iv) give exactly two radii throughout that range. Up to and including $A_2$ they are $\{0,A\}$; above $A_2$ both radii are positive. The split occurs at $10\log_{10}(A_2^2)\approx10.6814$~dB, just below the tabulated upper endpoint. For example, $A=3.2$ corresponds to $10.10$~dB, and the certificate on $3.198\le A\le3.202$ bounds the mass at the origin between $0.0809973$ and $0.0816880$~\cite{p2pcert}. The difference concerns the numerical count, not the finite-radius, uniform-phase characterization of~\cite{shamai1995}.

In the quadrature channel the three regimes correspond to familiar constellation shapes in the limit of many phases: a constant envelope (as for phase-shift keying), a constant envelope with an additional zero symbol, and two concentric rings (as for amplitude and phase-shift keying). With total noise power $\E\|Z\|^2=2$, the peak SNR is instead $10\log_{10}(A^2/2)$ in decibels, and the two thresholds are $4.79$ and $7.67$~dB. Both laws hold with the same structure as for the scalar channel (Table~\ref{tab:const}); the inner ring grows as $r^*\approx1.769\,(A-A_2)^{1/2}$, faster than the inner pair of the scalar channel ($1.318\,(A-A_2)^{1/2}$), while the constants $K_1$ and $K_2$ are smaller, so the leading rate loss at the same excess amplitude is smaller. In floating point, the next transition of the quadrature channel, where a mass point re-enters at the origin of the two-ring input, occurs at $A\approx4.72$.

\section{Discussion}
\label{sec:disc}

\emph{Structure of the transition.} Theorem~\ref{thm:main} is consistent with the local theory of~\cite{barletta2026}: the support changes at the origin, from three to four points, by the splitting of the origin into a symmetric pair. The four-point input is optimal throughout $(A_2,A_3]$ by Theorems~\ref{thm:main}(v) and~\ref{thm:third}(i). Theorem~\ref{thm:laws}(ii) describes the split locally: $\rho^*(A)/(A-A_2)\to c\in[1.7374087,1.7374097]$, so $r^*\approx1.3181\,(A-A_2)^{1/2}$, and the inner mass is continuous, $\mu^*(A_2+)=\lam^*(A_2)\approx0.27650$; the first certified box of Theorem~\ref{thm:main}(v) ($\rho\le6.1\times10^{-9}$ at $A_2+2.1\times10^{-9}$) agrees with this. At $A_2$ the zero of the KKT function at the origin has multiplicity exactly four in $x$, since $\Phi''_{\lam^*}<0$ near $s=0$ (the C$_0$ cells), as in the splitting mechanism of~\cite{barletta2026}; moreover, the transversality certificate shows that $g$ crosses zero with nonzero speed. At $A_3$, $\Phi_{P^*}(0)=C(A_3)$ but the four-point input still assigns zero mass to the origin (Fig.~\ref{fig:support}). The mass is positive on $(A_3,A_3^+]$ by Theorem~\ref{thm:third}(iii). Here $G_0$ crosses zero transversally, as $g$ does at $A_2$. At $A=2.786$, Corollary~\ref{cor:2786} certifies ternary optimality; Fig.~\ref{fig:kkt}(b) illustrates the corresponding information-density inequality.

\emph{Scope and trust.} Generating the certificates took about $131$ CPU hours for Theorem~\ref{thm:main}, $149$~h for the four-point cells of Theorem~\ref{thm:third} and $0.4$~h for its five-point cells, and $45$~h for Theorem~\ref{thm:quad}. The enclosures rely on Arb, the integrands in the archive~\cite{p2pcert}, and the tail bounds of Appendix~\ref{app:formulas}. Appendix~\ref{app:sound} proves the cell tests. Auditing the stored enclosures checks their signs and coverage without reevaluating the integrals. In a spot check of Theorem~\ref{thm:third}, we recomputed Krawczyk and sign tests on $37$ regular cells and the gap cell, sampling $1{,}178$ radial cells in total. This checks selected enclosures, not the integration software itself.

\section{Conclusion}

For the scalar channel, the ternary input is optimal on $(A_1,A_2]$ and the four-point input on $(A_2,A_3]$. The certified threshold $A_2\approx2.9074640719$ confirms the estimate $2.9075$ rather than $2.786$. A mass returns to the origin just above $A_3\approx4.02127532$. The first two quadrature transitions follow the same appearance and splitting laws. A new mass grows linearly in the excess amplitude, whereas a new inner radius grows as its square root. At all five transitions the rate lost on the previous stationary branch is quadratic with a positive coefficient. Capacity therefore has a continuous first derivative but unequal one-sided second derivatives.

It remains open whether every support transition has this regularity. A general proof would have to control the non-degeneracy conditions beyond the five certified thresholds. Do the same laws hold at later transitions or in higher dimensions? For the quadrature channel, a natural next case is the apparent return of mass to the origin beyond the certified two-ring interval.

\appendices
\section{Soundness of the Cell Certificates}
\label{app:sound}

\begin{proof}[Proof of Lemma~\ref{lem:sound}]
Fix $A\in[A_l,A_r]$, let $\lam=\lam^*(A)$ and $G=\Phi_\lam-\Phi_\lam(0)$, so that $G(0)=0$ and $G(A^2)=-\cH(\lam,A)=0$. \emph{Left part.} If $k_0>0$ and $g(A)=G'(0)\le0$, then $G''<0$ on $[0,t_{k_0}]$ gives $G'<0$ on $(0,t_{k_0}]$, hence $G\le0$ there. By induction on $k_0\le i<k$, $G(t_i)\le0$; on a D- or C-cell $G$ is decreasing (for C, $G'$ is decreasing and $G'(t_i)<0$), so $G\le G(t_i)\le0$, and on a V-cell $G<0$. \emph{Right part.} Let $j\ge k$ be the smallest index with $t_{j+1}>A_l^2$. The cells $j,\dots,m-1$ are of type I, so $G$ is increasing on $[t_j,A_r^2]$, and $t_j\le A_l^2\le A^2$ gives $G\le G(A^2)=0$ on $[t_j,A^2]$. For $k\le i<j$ we go backwards: $G(t_{i+1})\le0$ is known, an I-cell gives $G\le G(t_{i+1})\le0$, and a V-cell gives $G<0$. Hence $\Phi_\lam\le\Phi_\lam(0)=\Phi_\lam(A^2)=I(P_\lam)$ on $[0,A^2]$, and Lemma~\ref{lem:kkt} gives optimality. If $g(A)>0$, Proposition~\ref{prop:exit} excludes every ternary input. If $k_0=0$, the left part needs no assumption on $g$.
\end{proof}

\begin{lemma}\label{lem:four}
Let $(\mu^*,\rho^*)$ solve $f(\mu,\rho)=0$, $P^*=P_{\mu^*,\rho^*}$, and $G=\Phi_{P^*}-\Phi_{P^*}(A^2)$. Let $0\le t_L\le\rho^*\le t_R<A^2$, and suppose that $G''<0$ on $[t_L,t_R]$, that the cells covering $[0,t_L]$ are of type I or V, those covering $[t_R,t_k]$ of type D, C, or V, and those covering $[t_k,A^2]$ of type I or V, with type I for every cell that reaches beyond $A_l^2$. Then $P^*$ is capacity-achieving.
\end{lemma}
\begin{proof}
The first component of $f$ gives $G(\rho^*)=0$, so $I(P^*)=\mu^*\Phi_{P^*}(\rho^*)+(1-\mu^*)\Phi_{P^*}(A^2)=\Phi_{P^*}(A^2)$ and $G=\Phi_{P^*}-I(P^*)$. The second component gives $G'(\rho^*)=0$, and $G''<0$ on $[t_L,t_R]$ gives $G\le G(\rho^*)=0$ there. Backward induction from $t_L$ (an I-cell gives $G\le G(t_{i+1})$, a V-cell $G<0$) gives $G\le0$ on $[0,t_L]$; forward induction from $t_R$ as in the left part above gives $G\le0$ on $[t_R,t_k]$; and backward induction from $G(A^2)=0$ as in the right part gives $G\le0$ on $[t_k,A^2]$. Lemma~\ref{lem:kkt} applies.
\end{proof}

\begin{lemma}\label{lem:five}
Let $P$ be either a four-point input $P_{\mu^*,\rho^*}$ with $f(\mu^*,\rho^*)=0$ or a five-point input $P_{\nu,\mu^*,\rho^*}$, $\nu>0$, with $F(\nu,\mu^*,\rho^*)=0$, and let $G=\Phi_P-\Phi_P(A^2)$. Let $0\le s_0\le t_L\le\rho^*\le t_R<A^2$, and suppose that $G(0)\le0$, that $G''<0$ on $[t_L,t_R]$, that the cells covering $[0,s_0]$ are of type D, C, or V, those covering $[s_0,t_L]$ of type I or V, those covering $[t_R,t_k]$ of type D, C, or V, and those covering $[t_k,A^2]$ of type I or V, with type I for every cell that reaches beyond $A_l^2$. Then $P$ is capacity-achieving.
\end{lemma}
\begin{proof}
For the five-point input, $f_0=0$ gives $\Phi_P(0)=\Phi_P(A^2)$, so in both cases $I(P)=\Phi_P(A^2)$ and $G=\Phi_P-I(P)$, as in the proof of Lemma~\ref{lem:four}, which also gives $G\le0$ on $[t_L,t_R]$, on $[s_0,t_L]$ (backward induction from $t_L$), and on $[t_R,A^2]$. On $[0,s_0]$, forward induction from $G(0)\le0$, as in the left part of the proof of Lemma~\ref{lem:sound}, gives $G\le0$. Lemma~\ref{lem:kkt} applies.
\end{proof}

\section{Transversality Formulas and Truncation Bounds}
\label{app:formulas}
Let $a=A^2/2$, $\kappa=A^2/4$, $\eta=\frac12$, $e_\lambda=(1-\lam)e^{-a}$, $\mathsf F_k(x)={}_0F_1(;\eta+k;x)$, so that in the scalar case $\cosh(A\sqrt{2u})=\mathsf F_0(au)$, and $\Delta=\lam+e_\lambda\,\mathsf F_0(au)=M_\lam(2u)$. With $u\sim\mathrm{Gamma}(\eta+1,1)$ (that is, $Q_{3,0}/2$ in the scalar case), the quantities in~\eqref{eq:gprime} are $\Phi'_\lam(0)=\frac12-h$ with $h=\E[\frac{\kappa}{\eta}\,e_\lambda\,\mathsf F_1(au)/\Delta]$,
\begin{equation*}
\partial_\lam h=-\E\Big[\frac{\kappa}{\eta}\,\frac{e^{-a}\mathsf F_1}{\Delta^2}\Big],
\end{equation*}
\begin{multline*}
\partial_A h = A\,\E\biggl[
 \frac{(1-a)e_\lambda\mathsf F_1+\frac{a e_\lambda u}{\eta+1}\mathsf F_2}{2\eta\Delta}\\
 -\frac{a e_\lambda\mathsf F_1\bigl(\frac{e_\lambda u}{\eta}\mathsf F_1-e_\lambda\mathsf F_0\bigr)}{2\eta\Delta^2}
 \biggr],
\end{multline*}
all with argument $au$, and $\cK_\lam=\E_{u\sim\mathrm{Gamma}(\eta,1)}[(1-e^{-a}\mathsf F_0(au))^2/\Delta]$; $\partial_A\cH$ is obtained by differentiating~\eqref{eq:Phi} under the expectation. The integrands of $\partial_\lam h$, of $\partial_A h/A$, and of $\cK_\lam$ are bounded by $\frac{\kappa}{\eta} e^{a}(1-\lam)^{-2}$, $\frac{1+2a+2au/\eta}{2\eta}$, and $2e^{a+u/2}(1-\lam)^{-1}$, respectively. The bound for the integrand of $\partial_A h$ includes the additional factor $A$.

For clarity, let $\bar u=169$, $z=s/2$, and $\Pi_j(r)=\int_{\bar u}^\infty u^{\max(0,\lceil j\rceil)}e^{-ru}\,du$. Since $\bar u\ge1$, this also bounds the integral with exponent $j$. Set $B(r)=\Pi_{\eta}(r)+a\Pi_{\eta-1}(r)$. For noise variance one, $\eta=1/2$ (scalar) or $\eta=1$ (quadrature), and $0\le\lam<1$, let $\varepsilon_{\rm tr}[f]$ denote the absolute error due solely to omitting $u>\bar u$ in the integral for $f$. The bounds used for the zero--sphere mixture are
\begin{align*}
 \varepsilon_{\rm tr}[\cH] &\le \frac{e^a B(1/2)+B(1)}{\Gamma(\eta)},\\
 \varepsilon_{\rm tr}[\Phi'_\lam(s)] &\le \frac{\kappa e^z\Pi_{\eta}(1/2)}{\eta\,\Gamma(\eta+1)},\\
 \varepsilon_{\rm tr}[\Phi''_\lam(s)] &\le \frac{\kappa^2 e^z\Pi_{\eta+1}(1/2)}{\eta^2\Gamma(\eta+2)},\\
 \varepsilon_{\rm tr}[\cK_\lam] &\le \frac{2 e^a\Pi_{\eta-1}(1/2)}{(1-\lam)\Gamma(\eta)}.
\end{align*}
Indeed, $|\ln M_\lam(2u)|\le u+a$. The first two logarithmic derivatives in $q=2u$ have absolute values at most $\kappa/\eta$ and $(\kappa/\eta)^2$, respectively, by the hypergeometric derivative identity and $\mathsf F_{k+1}\le\mathsf F_k$ on the nonnegative axis. For shape $k\ge1/2$, the inequality $\,{}_0F_1(;k;zu)\le e^{2\sqrt{zu}}\le e^{u/2+2z}$ bounds the noncentral density by $e^{z-u/2}u^{k-1}/\Gamma(k)$. Integrating these bounds and the bound for the $\cK_\lam$ integrand above gives the four estimates.

For the four- and five-point inputs (and the two-ring input in dimension two), write $M=M_P(2u)=\sum_i w_i p_i$, with $p_i=e^{-a_i}\mathsf F_0(a_i u)$, $a_i=\rho_i/2\le a$, $w_i\ge0$, and $\sum_i w_i=1$. Since $e^{-a}\le p_i\le e^u$, the same bound on $|\ln M|$ holds. The ratios $M_q/M$ and $M_{qq}/M$ are convex combinations of the component ratios (subscripts denote derivatives in $q=2u$). These lie in $[0,\kappa/\eta]$ and $[0,\kappa^2/(\eta(\eta+1))]$, respectively. Thus the two radial derivative bounds above hold with $\Phi_P$ in place of $\Phi_\lam$. For the gap $d(s)=\Phi_P(s)-\Phi_P(0)$, integration against the noncentral and central densities gives
\[
 \varepsilon_{\rm tr}[d(s)]\le \frac{e^z B(1/2)+B(1)}{\Gamma(\eta)} .
\]
At $s=A^2$ this recovers the first bound for the zero--sphere mixture, since $\cH=-d(A^2)$. The bound for $\cK_\lam$ concerns that mixture's weight derivative; it is not a bound on the effective curvature at the third transition.

For the Jacobian bounds, let a dot denote a parameter derivative at fixed $s$. Choose $h_i\ge p_i/M$ for each varying weight: $h_i=1/w_i$ suffices when $w_i>0$. For the origin atom, $p_i=1$, so $h_i=e^{a_j}/w_j$ is valid even at $\nu=0$ whenever another atom has $w_j>0$, as in Section~\ref{sec:cert3}. Use upper bounds uniform on the parameter cell. Put $W=\sum_{\dot w_i\ne0}|\dot w_i|h_i$ and $R=\tfrac12\sum_i|\dot\rho_i|$. Differentiating the component kernels and using $\mathsf F_{k+1}\le\mathsf F_k$ gives $|\dot M/M|\le U(u)=W+R(1+u/\eta)$. The quotient rule then gives $|\partial_q(\dot M/M)|\le V(u)$, where
\[
 V(u) = \frac{\kappa}{\eta}\bigl(W+U(u)\bigr)
 + \sum_i\frac{|\dot\rho_i|}{4\eta}\left(1+a_i+\frac{a_i u}{\eta+1}\right).
\]
For an affine envelope $L(u)=L_0+L_1u$ with nonnegative coefficients write $\mathcal T_k(L)=[L_0\Pi_{k-1}(1/2)+L_1\Pi_k(1/2)]/\Gamma(k)$. Integration yields
\begin{align*}
 \varepsilon_{\rm tr}[\dot d(s)]&\le(e^z+1)\mathcal T_\eta(U),\\
 \varepsilon_{\rm tr}[\dot\Phi'_P(s)]&\le e^z\mathcal T_{\eta+1}(V).
\end{align*}
When the evaluation point $s$ also varies, the chain rule adds $\Phi'_P(s)\dot s$ and $\Phi''_P(s)\dot s$, respectively, with the radial bounds above.

\end{document}